\documentclass{moriond}
\usepackage{graphicx}
\usepackage{sidecap}

\def\BelleII{\emph{Belle II }}
\def\be{\begin{equation}}
\def\ee{\end{equation}}
\def\bea{\begin{eqnarray}}
\def\eea{\end{eqnarray}}

\newcommand{\Photo}{}

\begin{document}
\vspace*{4cm}
\title{LATEST BELLE II RESULTS ON BEAUTY AND CHARM DECAYS}

\author{S. SANDILYA \\ (on behalf of the \BelleII Collaboration)}

\address{Department of Physics, Indian Institute of Technology Hyderabad,\\
Telangana, 502285, India}

\maketitle\abstracts{
 We present the measurements, performed by the Belle II experiment, related to the $B$ and $D$ meson decays. These results are based on 63 ${\rm fb^{-1}}$ and  9 ${\rm fb^{-1}} $ of $e^{+}e^{-}$ collision data recorded by the Belle II detector at a center-of-mass energy corresponding to the mass of the $\Upsilon (4S)$ resonance and 60 MeV below the $\Upsilon (4S)$ resonance. The results reassure that Belle II is in the right direction in pursuit of measuring the Standard Model predictions with improved precision.}

\section{Introduction}
The prime objective of the Belle II experiment is to measure the parameters of Standard Model (SM) with better precision and to search for the signs of new physics (NP) beyond the SM~\cite{B2TiP}.
The SuperKEKB asymmetric-energy $e^{+}e^{-}$ collider facilitates large collision rate with its designed instantaneous luminosity $6.5\times 10^{35}{\rm ~cm^{2}s^{-1}}$~\cite{SuperKEKB}.
The Belle II detector ~\cite{belle2} consists of several sub-detector components located around the interaction region of SuperKEKB in a cylindrical geometry. The $e^{+}$ and $e^{-}$ beams collide at a center-of-mass (CM) energy equal to the mass of the $\Upsilon{\rm (4S)}$ resonance ($\sqrt{s} = 10.58$ GeV), which leads to a clean sample of quantum-correlated pairs of $B$ mesons. Apart from $\Upsilon{\rm (4S)}$, charm quark pairs are also produced in the $e^{+}e^{-}$ collisions with a similar cross-section, generating equivalently large samples of $D$ mesons. Here, we report the measurements with the $B$ and $D$ meson decays based on 63 ${\rm fb^{-1}}$ data recorded at the $\Upsilon{\rm (4S)}$ and an additional 9  ${\rm fb^{-1}}$ recorded 60 MeV below the $\Upsilon{\rm (4S)}$ resonance peak. (Inclusion of charge-conjugate processes is implied throughout this article.)

\section{Study of {\boldmath $B\to D^{(*)}h$ ($h=\pi , K$)} decays}
The decays $B^{-}\to D^{(*)0}K^{-}$ arise from the $b\to c\bar{u}s$ quark-level transition and contribute to the measurement of the CKM angle $\phi_{3}$ (or $\gamma$)~\cite{phi3}. And, the decays $\bar{B}\to D^{(*)}\pi^{-}$ are amongst the dominant $B$ decay modes and serve as a good control sample for the reconstruction procedure. The decay modes reconstructed are : (1) $B^{-}\to D^0 h^-$, $D^{0}\to K^-\pi^+$ or $K^{0}_{\rm S}\pi^+\pi^-$; (2) $B^{-}\to D^{*0}h^-$, $D^{*0}\to D^{0}\pi^0$, $D^{0}\to K^{-}\pi^+$; (3) $\overline{B}^{0}\to D^+ h^-$, $D^+\to K^-\pi^+\pi^+$; and (4) $\overline{B}^{0}\to D^{*+} h^-$, $D^{*+}\to D^0\pi^+$, $D^0\to K^-\pi^+$. The final-state particles ($\pi^{+}, K^{+}, K^{0}_{S}$ and $\pi^{0}$) are selected and combined to form a $D^{(*)}$ meson candidate. A $B$ meson candidate is formed by combining a $D^{(*)}$ candidate and a prompt $h$ candidate selected with a requirement on hadron identification either as a kaon or as a pion. In order to discriminate the signal from background, two kinematic variables are introduced: the beam constrained mass $M_{\rm bc} \equiv \sqrt{E^2_{\rm beam} - \left({\vec{p}}_{B}\right)^2}$  and the energy difference $\Delta E \equiv E_{B} - E_{\rm beam}$, where $E_{\rm beam}$ is the beam energy and $E_{B}$ and ${\vec{p}}_{B}$ are the energy and momentum, respectively, of the reconstructed $B$ candidate. All these quantities are calculated in the $e^{+}e^{-}$ CM frame. Signal $B$ candidates are selected by applying criterion on $M_{\rm bc}$, and the final signal yield is extracted by performing a maximum-likelihood (ML) fit to $\Delta E$. The ratio between the decays $B^{-}\to D^{(*)0}K^{-}$/$B^{-}\to D^{(*)0}\pi^{-}$  have been reported~\cite{BtoDK}, which are found to be compatible with the world-average values.

\section{Study of hadronic {\boldmath $B$} decays to charmless mesons}

\subsection{Reconstruction of $B^{0}\to\pi^{0}\pi^{0}$}
Precise measurements of each $B\to\pi\pi$ decay is crucial to invoke the isospin sum-rule to disentangle the shift in the value of CKM angle $\phi_{2}$ due to the presence of gluonic penguin~\cite{isospin1}. The decay $B^{0}\to\pi^{0}\pi^{0}$ is of particular interest at Belle II, as the final state consists of four photons. The signal yield is extracted from a simultaneous unbinned ML fit to $M_{\rm bc}$, $\Delta E$ and the output of a fast boosted decision tree (FBDT) algorithm based on event topological variables. The procedure is validated using the control sample of $B^{0}\to D^{0}[K^-\pi^+\pi^0]\pi^{0}$. The signal yield is $[14.0^{+6.8}_{-5.6}]$, corresponding to a branching fraction ${\cal B}(B^{0}\to\pi^{0}\pi^{0}) = (0.98^{+0.48}_{-0.39} ~\pm 0.27) \times 10^{-6}$. This result agrees with the previous measurements.

\subsection{Analysis of $B^{0}\to\rho^{+}\rho^{0}$}
The decay $B^{0}\to\rho^{+}\rho^{0}$ has two vector mesons in the final state, and is dominated by their longitudinal polarization components. The large width of the $\rho$ meson and neutral pions in the final state lead to large background contamination. The signal yield is determined from an unbinned ML fit to $\Delta E$, FBDT output, reconstructed invariant masses of $\rho^{+}$ and $\rho^{0}$, and their helicity angles. We find $104\pm 16$ signal events which correspond to ${\cal B}(B^{0}\to\rho^{+}\rho^{0}) = (20.6\pm 3.2 \pm 4.0) \times 10^{-6}$, and the fraction of longitudinal polarization is $[0.936^{+0.049}_{-0.041} \pm 0.021]$.

\subsection{Analysis of $B^{0}\to K^{0}\pi^{0}$}
The $K\pi$ isospin sum rule~\cite{isospin2} offers a stringent null test of the SM, and is expressed in terms of direct $CP$ asymmetries and {$\cal B$} of the four $B\to K\pi$ decay modes. The experimental uncertainty of the sum rule is currently dominated by the measurements of the decay $B^{0}\to K^{0}\pi^{0}$. We obtain $45^{+9}_{-8}$ signal events for this decay by fitting $\Delta E$ and $M_{\rm bc}$ distributions, which is translated to ${\cal B}(B^{0}\to K^{0}\pi^{0}) = (8.5^{+1.7}_{-1.6} \pm 1.2) \times 10^{-6}$.~\cite{BtoK0pi0}. As this decay is a $CP$ eigen-state, based on the output of flavor tagger~\cite{flavortag}, we find the decay-time-integrated direct $CP$ asymmetry as $[-0.40^{+0.46}_{-0.44} \pm 0.04]$.

\section{Reconstruction of the decay {\boldmath $B^{0} \to J/\psi K^{0}_{L}$}}
The decay $B^{0} \to J/\psi K^{0}_{L}$ provides an independent measurement of the CKM angle $\sin (2\phi_{1})$. The $J/\psi$ candidates are reconstructed from both $\mu^{+}\mu^{-}$ and $e^{+}e^{-}$ final states. The $K^{0}_{L}$ candidate is identified from hadronic shower cluster in the $K^{0}_{L}$ and muon sub-detector. The energy and momentum of the $K^{0}_{L}$ is inferred with the help of flight direction while reconstructing $B^{0} \to J/\psi K^{0}_{L}$. The signal yield is extracted by performing a fit to the $\Delta E$ distribution. We find $267\pm 21 {\rm (stat.)}$ and $226\pm 20 {\rm (stat.)}$, signal events in $\mu^{+}\mu^{-}$ and $e^{+}e^{-}$ final states, respectively. The signal yields are consistent with that from the previous Belle experiment with a similar purity. In future, we plan to improve the signal yield by also including the hadronic clusters in the electromagnetic calorimeter.

\section{Radiative and electroweak penguin {\boldmath $B$} decays}
The $B$ decays mediated by flavour-changing-neutral-current transitions $b\to s \gamma$,  $b\to s \ell^{+}\ell^{-}$ and $b\to s \nu\bar{\nu}$, proceed through penguin loop and box diagrams. These decays are considered excellent probes for the NP.   
\subsection{Photon energy spectrum of the inclusive decay $B\to X_{s}\gamma$}
\begin{SCfigure}
  \includegraphics[width=0.5\linewidth]{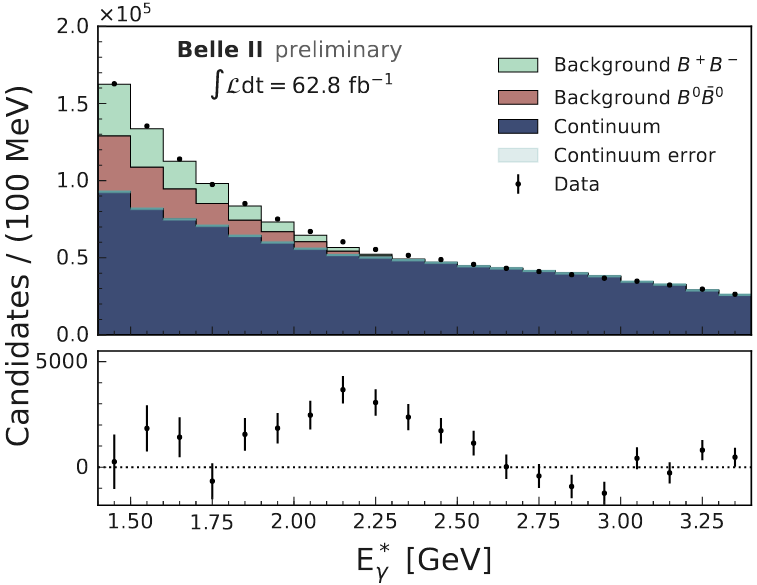}
  \caption[]{(Top) The photon energy spectrum of the selected $b\to s\gamma$ candidates in the CM frame. The data events are represented by black dots and the expected background events are represented by various color-filled histograms. (Bottom) The background subtracted distribution of data events indicates the presence of $b\to s\gamma$ events.}
  \label{fig:btosg}
\end{SCfigure}

Moments of the photon energy spectrum in the $b\to s \gamma$ transition are important for the determination of $b$--quark mass and its motion inside the $B$ meson. In this analysis~\cite{btosg}, we first search for a high-energy photon ($E^{*}_{\gamma}> 1.4$ GeV in the CM frame) in the collision events, which should not be arising from a $\pi^{0}$ or an $\eta$ meson decay. The expected spectrum of background events in data is obtained by scaling the Monte Carlo distribution fromthe off-resonance sample and the sidebands. A clear visible excess in data is observed around the expected region of $E^{*}_{\gamma}$ as shown in figure~\ref{fig:btosg}.
\subsection{$B^{+} \to K^{+}\ell^{+}\ell^{-}$ in the early data sample of Belle II}
The decays $B^{+}\to K^{+}\ell^{+}\ell^{-}$ have raised a lot of interest in study of lepton-flavor-universality ratios. These rare decays ( ${\cal B}\sim 10^{-7}$) are challenging to be observed in the early data sample of Belle II. In a recent analysis at Belle II, $8.6^{+4.3}_{-3.9}\pm 0.4$ signal events are observed for the decay $B^{+}\to K^{+}\ell^{+}\ell^{-}$~\cite{BtoKll}.

\subsection{Search for the decay $B^{+} \to K^{+}\nu\bar{\nu}$}
The decay $B^{+} \to K^{+}\nu\bar{\nu}$ is theoretically cleaner as there is no contribution from the virtual photon in the diagrams unlike the decays mediated by the $b \to s\ell^{+}\ell^{-}$ transition. The decay  $B^{+} \to K^{+}\nu\bar{\nu}$ is experimentally challenging as the final state consists of two neutrinos. At Belle II, a new analysis method has been developed for this search that exploits the topological features of the decay leading to a larger signal reconstruction efficiency.  The method has been validated with $B^{+} \to K^{+}J/\psi[\to \mu^{+}\mu^{-}]$, by ignoring the  $\mu^{+}\mu^{-}$ from the $J/\psi$ decay to mimic the missing energy from two neutrinos; $K^{+}$ momentum is also modified to correspond to a three-body decay. The signal strength obtained from the fit is $4.2^{+2.9}_{-2.8} {~\rm (stat.) ^{+1.8}_{-1.6}} {~\rm (syst.)}$, which corresponds to ${\cal B} = [1.9 \pm 1.3 {~\rm (stat.) ^{+0.8}_{-0.7}} {~\rm (syst.)}] \times 10^{-5}$. This measurement is competitive with the previous results, taking into account of the size of data sample used. As no significant signal is observed, the expected 90\% confidence-level (CL) upper limit (UL) on the branching fraction of $2.3\times 10^{-5}$ is derived and the observed UL of $4.1\times 10^{-5}$ is set at 90\% CL.   
\section{Study of charm meson decays}
\subsection{Measurement of the $D^{0}$ lifetime}
The lifetime measurement of $D^{0}$ meson is performed using $9.6{~\rm fb^{-1}}$ Belle II data recorded in 2019~\cite{D0Life}. Coming from the decay $D^{*+} \to D^{0}\pi^{+}$ in charm-quark pair events, the $D^{0}$ meson is reconstructed in three decay modes: $D^{0}\to K^{-}\pi^{+}$, $D^{0}\to K^{-}\pi^{+}\pi^{0}$ and $D^{0}\to K^{-}\pi^{+}\pi^{+}\pi^{-}$. The $D^{0}$ lifetime is obtained by performing a two-dimensional unbinned ML fit to distributions of proper time and its uncertainty. The average lifetime of $D^{0}$ meson is measured to be $412.3\pm 2.0 {~\rm fs}$. With $72 {~\rm fb^{-1}}$ of Belle II data, the lifetime measurement of $D^{0}$ is expected to be competitive with the world-averages.

\subsection{Preliminary analysis of {$D^{*+} \to D^{0}[\to \pi^{+}\pi^{-}\pi^{0}]\pi^{+}$}}
$CP$ violation in the charm sector is an important topic to study at the Belle II experiment. The ultimate aim is to perform a time-averaged Dalitz plot analysis for the decay $D^{*+} \to D^{0}[\to \pi^{+}\pi^{-}\pi^{0}]\pi^{+}$. Currently, the decay has been reconstructed and the signal yield is extracted using a binned ML fit to the distribution of $\Delta M$, difference between the reconstructed invariant masses of $D^{*+}$ and $D^{0}$. The signal yield per ${\rm fb^{-1}}$ is found to be $305\pm 15 {~\rm (stat.)}$~\cite{DeltaM}.   

\section*{Summary}
Belle II has been recording data steadily despite the CoViD-19 pandemic towards its ultimate goal of accumulating at least $50 {~\rm ab^{-1}}$ of $e^{+}e^{-}$ collision data. This upcoming large and clean samples of $B$ and $D$ mesons (and $\tau$ leptons) will allow Belle II to search for NP and improve the measurements of various SM parameters. Measurements reported in this article are based on the early data recorded with the Belle II detector and some of them analysed using novel methods. These results demonstrate that all the subdetectors of Belle II are performing as per expectation and some results are already getting competitive.

\end{document}